\documentclass[preprint]{revtex4}
\usepackage{epsfig}

\def\beq{\begin{equation}}
\def\eeq{\end{equation}}
\def\beqa{\begin{eqnarray}}
\def\eeqa{\end{eqnarray}}

\begin{document}

\title{\sc Korteweg - de Vries  solitons in relativistic hydrodynamics}

\author{D.A. Foga\c{c}a\dag\  and F.S. Navarra\dag\ }
%\affiliation{Instituto de F\'{\i}sica, Universidade de S\~{a}o Paulo\\
\address{\dag\ Instituto de F\'{\i}sica, Universidade de S\~{a}o Paulo\\
 C.P. 66318,  05315-970 S\~{a}o Paulo, SP, Brazil}

\begin{abstract}

In a previous work, assuming that the nucleus can be treated as a perfect fluid,  we have 
studied the propagation of perturbations in the baryon density. For a given equation of state 
we have derived a  Korteweg - de Vries (KdV) equation from Euler 
and continuity equations in non-relativistic  hydrodynamics.  Here, using a more general   
equation of state, we extend our formalism to  relativistic hydrodynamics.  

\end{abstract} 

%\pacs{PACS Numbers~ :~ 12.38.Lg, 12.40.Yx, 12.39.Mk}
\maketitle

%%%%%%%%%%%%%%%%%%%%%%%%%%%%%%%%%%%%%%%%%%%%%%%%%%%%%%%%%%%%%%%%%%%%%%%%%
% Beginning of the paper
%%%%%%%%%%%%%%%%%%%%%%%%%%%%%%%%%%%%%%%%%%%%%%%%%%%%%%%%%%%%%%%%%%%%%%%%%

%%%%%%%%%%%%%%%%%%%  Introduction %%%%%%%%%%%%%%%%%%%%%%%%%%%%%%%%%%%%%%%

\vspace{1cm}
\section{Introduction}

Long ago \cite{frsw} it was suggested that Korteweg - de Vries solitons might be formed 
in the  nuclear medium. In a  previous work \cite{nois}  we have updated the early works 
on the subject introducing a realistic equation of state (EOS) for nuclear matter.  We have 
found that these solitary waves can indeed  exist in the nuclear medium, provided that 
derivative couplings between the nucleon and the vector field are included. These couplings 
lead to an 
energy density which depends on the  Laplacian of the baryon density. For this class of 
equations  of state, which is quite general (as pointed out in \cite{furn,fst97}), 
perturbations on the nuclear density can propagate as a pulse without dissipation.  

During the analysis of 
several realistic nuclear equations of state, we realized that, very often the speed of sound 
$c_s$ is in the range $0.15 -0.25$. Compared to the speed of light these values are not large 
but not very small either. This suggests that, even for slowly moving nuclear matter, 
relativistic effects might be sizeable. This concern motivates  the extension of the 
formalism presented in \cite{nois} to relativistic hydrodynamics.

In the next section we review the most relevant  equations  writting them in an 
appropriate form for the subsequent manipulations. In the following section we discuss three 
models for  the equation of state and in the next section we derive the KdV equation for the 
proposed models and  present their solutions. In the final section we present 
our conclusions.

%\hspace{0.2cm}

\section{Relativistic hydroynamics}

In this section we review the main expressions of  relativistic 
hydrodynamics. In natural units ($c=1$) the velocity four vector  $u^{\nu}$ is defined as:
\begin{equation}
u^{\nu} \,\, = \,\, (u^{0},\vec{u}) \,\,  = \left(\gamma , \gamma \vec{v} \right)
\label{us}
\end{equation}
where $\gamma$ is the Lorentz contraction factor given by:
\begin{equation}
\gamma=(1-v^{2})^{-1/2}
\label{lorentz}
\end{equation}
The velocity field of  matter is $\vec{v}=\vec{v}(t,x,y,z)$
and thus $u^{\nu}u_{\nu}=1$. The energy-momentum tensor is, as usual, given by:
\begin{equation}
T_{\mu \nu}=(\varepsilon +p)u_{\mu}u_{\nu}-pg_{\mu\nu}
\label{tensor}
\end{equation}
where  $\varepsilon$ and $p$ are the energy density and pressure respectively. 
Energy-momentum conservation is  ensured by:
\begin{equation}
\partial_{\nu}{T_{\mu}}^{\nu}=0
\label{cons}
\end{equation}
The projection of (\ref{cons}) onto  a direction perpendicular to $u^{\mu}$ gives us
the relativistic version of  Euler equation \cite{wein,land,elze}:
\begin{equation}
{\frac{\partial {\vec{v}}}{\partial t}}+(\vec{v} \cdot \vec{\nabla})\vec{v}=
-{\frac{1}{(\varepsilon + p)\gamma^{2}}}
\bigg({\vec{\nabla} p +\vec{v} {\frac{\partial p}{\partial t}}}\bigg)
\label{eul}
\end{equation}
The relativistic version of the continuity equation for the baryon number is  
\cite{wein,land,elze}:
\begin{equation}
\partial_{\nu}{j_{B}}^{\nu}=0
\label{conucleon}
\end{equation}
Since ${j_{B}}^{\nu}=u^{\nu} \rho_{B}$ the above equation reads
\begin{equation}
{\frac{\partial}{\partial t}}(\rho_{B}\gamma)+\vec{\nabla} \cdot (\rho_{B}\gamma \vec{v})=0
\label{con}
\end{equation}
The enthalpy per nucleon is given by \cite{land}:
\begin{equation}
dh=Tds+Vdp 
\label{ent1}
\end{equation}
where \hspace{0.2cm} $V=1/\rho_{B}$ \hspace{0.2cm} is the specific volume. 
For a perfect fluid  $(ds=0)$    the  equation above becomes  $dp=\rho_{B}dh$ and 
consequently:
\begin{equation}
\vec{\nabla} p=\rho_{B}\vec{\nabla} h, \hspace{1cm} 
{\frac{\partial p}{\partial t}}=\rho_{B}{\frac{\partial h}{\partial t}}
\label{deripe}
\end{equation}
Inserting (\ref{deripe}) in (\ref{eul}) we find:
\begin{equation}
{\frac{\partial {\vec{v}}}{\partial t}}+(\vec{v} \cdot \vec{\nabla})\vec{v}=
-{\frac{\rho_{B}}{(\varepsilon + p)\gamma^{2}}}
\bigg({\vec{\nabla} h +\vec{v} {\frac{\partial h}{\partial t}}}\bigg)
\label{euler}
\end{equation}
Recalling the Gibbs  relation \cite{reif}:
\begin{equation}
\varepsilon - Ts + p = \mu_{B} \rho_{B} 
\label{gibbs}
\end{equation}
and considering the case where $T=0$ we obtain:
\begin{equation}
\varepsilon + p=\mu_{B} \rho_{B}
\label{gibbsrel}
\end{equation}
where $\varepsilon$, $p$,  $\mu_{B}$  and  $\rho_{B}$ are the energy density, pressure,  
baryochemical potential and  baryon density respectively.
Inserting (\ref{lorentz}) and  (\ref{gibbsrel}) into 
(\ref{euler}) we  obtain:
\begin{equation}
{\frac{\partial {\vec{v}}}{\partial t}}+(\vec{v} \cdot \vec{\nabla})\vec{v}= \, - \, 
{\frac{(1 - v^{2})}{\mu_{B}}}
\bigg({\vec{\nabla} h +\vec{v} {\frac{\partial h}{\partial t}}}\bigg)
\label{eulerfinal}
\end{equation}

We close this section comparing the relativistic and 
non-relativistic versions of the Euler and continuity equations. The latter were 
presented in \cite{nois}:
\begin{equation}
{\frac{\partial \rho_B}{\partial t}} + {\vec{\nabla}} \cdot (\rho_B {\vec{v}})=0  
\label{contibari}
\end{equation} 
\begin{equation}
{\frac{\partial \vec{v}}{\partial t}} +(\vec{v} \cdot \vec{\nabla}) \vec{v}=
-\bigg({\frac{1}{M}}\bigg) \vec{\nabla} h 
\label{eulerentalpia}
\end{equation}
and the former are (\ref{con}) and   (\ref{eulerfinal}):
\begin{equation}
{\frac{\partial}{\partial t}}(\rho_{B}\gamma)+\vec{\nabla} \cdot (\rho_{B}\gamma \vec{v})=0
\label{conf}
\end{equation}
\begin{equation}
{\frac{\partial {\vec{v}}}{\partial t}}+(\vec{v} \cdot \vec{\nabla})\vec{v}= \, - \, 
{\frac{(1 - v^{2})}{\mu_{B}}}
\bigg({\vec{\nabla} h +\vec{v} {\frac{\partial h}{\partial t}}}\bigg)
\label{eulerfinalf}
\end{equation}
The two pairs are similar. The differences are only in the $\gamma$ factors and in the  
last term of (\ref{eulerfinalf}), where the appearance of  time derivative reflects  the 
symmetry between space and time. Since the enthalpy per nucleon may also be written 
as \cite{nois,abu}:
\begin{equation}
h=\frac{\partial \varepsilon}{\partial \rho_{B}} 
\label{entalbu}
\end{equation} 
it becomes clear that the ``force'' on the 
right hand side of  ({\ref{eulerentalpia}}) and (\ref{eulerfinalf}) will be ultimately  
determined by the equation of state, i.e., by the function $\varepsilon(\rho_B)$.

\section{Equation of state}

Equations  ({\ref{eulerentalpia}}) and (\ref{eulerfinalf})  contain the gradient of the 
derivative of the energy density. If $\varepsilon$ contains a 
Laplacian of $\rho_B$, i.e., ${\mathcal{\varepsilon}} 
\propto ... + ... \nabla^{2} \rho_{B} + ...$, then 
({\ref{eulerentalpia}}) and (\ref{eulerfinalf}) will have a cubic derivative
with respect to the space coordinate, which will give rise to the Korteweg-de Vries equation 
for the baryon density. 
The most popular  relativistic mean field models do not have higher derivative terms and, 
even if they have at the start, these terms are usually neglected during the calculations.  

In \cite{nois} we have added a new derivative term to  the usual non-linear QHD 
\cite{lala}, given by
\begin{equation}
{\mathcal{L_{M}}} \equiv {\frac{g_{v}}{{m_{v}}^{2}}}\bar{\psi}
(\partial_{\nu} \partial^{\nu} V_{\mu})\gamma^{\mu} \psi 
\label{lagram}
\end{equation}
where, as usual, the degrees of freedom are 
the baryon field $\psi$, the neutral scalar meson field $\phi$
and the neutral vector meson field $V_{\mu}$, with the respective couplings and masses. 
The new term is designed to be small in comparison with the main baryon - vector meson 
interaction term $g_{v} \bar{\psi} \gamma_{\mu} V^{\mu}  \psi$. 
Folowing the standard steps of the mean field formalism we have arrived at the following
expression for the energy density \cite{nois}:
\begin{eqnarray}
\varepsilon&=&{\frac{{g_{v}}^{2}}{2{m_{v}}^{2}}}\rho_{B}^{2} 
+{\frac{{m_{s}}^{2}}{2}}{\bigg[{\frac{(M^{*}-M)}{g_{s}}}\bigg]}^{2}
+{\frac{\eta}{(2\pi)^{3}}}\int_{0}^{k_{F}} d^3{k} ({\vec{k}}^{2}+{M^{*}}^{2})^{1/2}  
+{\frac{b}{3g_s^3}}(M^{*}-M)^{3} \nonumber \\
&+&{\frac{c}{4g_{s}^{4}}}(M^{*}-M)^{4} + {\frac{{g_{v}}^{2}}{{m_{v}}^{4}}}\rho_{B}
\nabla^{2}\rho_{B}
\label{epsilonexp}
\end{eqnarray}
where $\eta$ is  the baryon  spin-isospin degeneracy factor,    
$M^*$ stands for the nucleon effective mass (given by $M^{*} \equiv M-g_{s}\phi_{0}$)  
and the constants $b$, $c$,  $g_s$ and $g_v$ were taken from \cite{lala}.  
Although Eq.  (\ref{epsilonexp}) 
was obtained with the help of a specific Lagrangian taken from \cite{lala} and a prototype 
Laplacian interaction (\ref{lagram}), the above  form of the energy density 
follows quite naturally from an approach based on the density functional theory 
\cite{fst97}, for a wide variety of  underlying Lagrangians.

We now follow the treatment developed in \cite{nois,frsw,abu} to 
obtain the Korteweg-de Vries  equation in one dimension through 
the combination of ({\ref{conf}}) and (\ref{eulerfinalf}). 
With the help of (\ref{epsilonexp}) we first calculate
the energy per nucleon given by $E=\varepsilon/\rho_{B}$. We next perform a  
Taylor expansion of $E$ around  the equilibrium density $\rho_{0}$ up to second order:
\begin{equation}
E(\rho_{B})=E(\rho_{0})
+\frac{1}{2}  
\bigg({\frac{\partial^{2} E}{{\partial\rho_{B}}^{2}}}\bigg)_{\rho_{B}=\rho_{0}} 
(\rho_{B}-\rho_{0})^{2} 
\label{energypernucleon}
\end{equation}
where the first order term vanishes because of the  saturation condition:
\begin{equation}
{\frac{\partial }{{\partial\rho_{B}}}}\bigg({\frac{\varepsilon}
{\rho_{B}}}-M\bigg)_{\rho_{B}=\rho_{0}}\hspace{0.25cm}=
\hspace{0.25cm}\bigg({\frac{\partial E}{{\partial\rho_{B}}}}\bigg)_{\rho_{B}=\rho_{0}}
\hspace{0.25cm}=\hspace{0.5cm}0
\label{satcond}
\end{equation}
We arrive at (for more details see \cite{nois}):
\begin{equation}
E(\rho_{B})= E(\rho_{0}) \,\,  
%\bigg({\frac{{g_{v}}^{2}}{2{m_{v}}^{2}}}\bigg){\rho_{0}}
%+ {\frac{\eta(\rho_{0})}{\rho_{0}}}
+\bigg({\frac{{g_{v}}^{2}}{{m_{v}}^{4}}}\bigg)(\nabla^{2}\rho_{B}) 
+\frac{1}{2}  \frac{M{c_{s}}^{2}}{\rho_{0}^{2}} (\rho_{B}-\rho_{0})^{2} 
\label{energypernucleoncaseI}
\end{equation}
The enthalpy per nucleon may also be written as \cite{abu}:
\begin{equation}
h=E+\rho_{B}{\frac{\partial E}{{\partial\rho_{B}}}}
\label{abuent}
\end{equation}
Using (\ref{energypernucleoncaseI}) to evaluate (\ref{abuent}) 
and its derivatives we find:
\begin{equation}
\vec{\nabla}h={\frac{3M{c_{s}}^{2}}{{\rho_{0}}^{2}}}\rho_{B}\vec{\nabla} \rho_{B}-
{\frac{2M{c_{s}}^{2}}{{\rho_{0}}}}\vec{\nabla}\rho_{B}+
{\frac{{g_{v}}^{2}}{{m_{v}}^{4}}}\vec{\nabla}(\vec{\nabla}^{2}\rho_{B})
\label{gradhmqhd}
\end{equation}
and
\begin{equation}
{\frac{\partial h}{\partial t}}={\frac{3M{c_{s}}^{2}}{{\rho_{0}}^{2}}}\rho_{B}
{\frac{\partial \rho_{B}}{\partial t}}-
{\frac{2M{c_{s}}^{2}}{{\rho_{0}}}}{\frac{\partial \rho_{B}}{\partial t}}+
{\frac{{g_{v}}^{2}}{{m_{v}}^{4}}}{\frac{\partial }{\partial t}}(\vec{\nabla}^{2}\rho_{B})
\label{htempmqhd}
\end{equation}

The expressions (\ref{gradhmqhd}) and (\ref{htempmqhd}), from now on referred to as model I,
will be inserted into (\ref{eulerfinalf}) as it will be seen in the next section.

We shall  now  consider a  more general expression for the energy density
\begin{equation}
\varepsilon= \alpha_{1}{\rho_{B}}+\alpha_{2}{{\rho_{B}}^{2}}+
\alpha_{3}{{\rho_{B}}^{3}} + \beta\rho_{B}\vec{\nabla}^{2}\rho_{B}
\label{ansatz}
\end{equation}
where $\alpha_i$ and $\beta$ are constants.  
This Ansatz is similar to the energy density  used in \cite{nois,frsw,abu} and is  
consistent with the EOS obtained with the  approach based on the density functional 
theory \cite{furn,fst97}. Let's assume that (\ref{ansatz}) is an appropriate model for 
nuclear matter and that it satisifes the  saturation condition (\ref{satcond}). This 
will be our model II. Once  again we  calculate
the energy per nucleon $E=\varepsilon/\rho_{B}$ then Taylor
expand  it around the equilibrium density $\rho_{0}$ up to second order 
(\ref{energypernucleon})  and  find
\begin{equation}
E(\rho_{B})=\alpha_{1}+\alpha_{2}{{\rho_{0}}}+
2\alpha_{3}{{\rho_{0}}^{2}} + \alpha_{3}{{\rho_{B}}^{2}}-2\alpha_{3}\rho_{B}\rho_{0}
+\beta\vec{\nabla}^{2}\rho_{B}
\label{energypernucleoncaseII}
\end{equation}
Now, using (\ref{energypernucleoncaseII}) to evaluate (\ref{abuent}) 
and its derivatives we find:
\begin{equation}
\vec{\nabla}h=6 \alpha_{3}\rho_{B}\vec{\nabla} \rho_{B}
-4 \alpha_{3}\rho_{0}\vec{\nabla}\rho_{B}+
\beta\vec{\nabla}(\vec{\nabla}^{2}\rho_{B})
\label{gradhcaseII}
\end{equation}
and
\begin{equation}
{\frac{\partial h}{\partial t}}=6 \alpha_{3}\rho_{B}{\frac{\partial \rho_{B}}{\partial t}}
-4 \alpha_{3}\rho_{0}{\frac{\partial \rho_{B}}{\partial t}}+
\beta{\frac{\partial }{\partial t}}(\vec{\nabla}^{2}\rho_{B})
\label{htempcaseII}
\end{equation}

In model III we consider  hadronic matter at arbitrary  constant baryon  density, but now 
no saturation condition is imposed.  This last choice is motivated by a future study of 
dense stars.  In this case we calculate the enthalpy directly from (\ref{ansatz}) and 
(\ref{entalbu}) obtaining the following expressions for the derivatives:
\begin{equation}
\vec{\nabla}h=6 \alpha_{3}\rho_{B}\vec{\nabla} \rho_{B}+
2 \alpha_{2}\vec{\nabla}\rho_{B}+
\beta\vec{\nabla}(\vec{\nabla}^{2}\rho_{B})
\label{gradhcaseIII}
\end{equation}
and
\begin{equation}
{\frac{\partial h}{\partial t}}=6 \alpha_{3}\rho_{B}{\frac{\partial \rho_{B}}{\partial t}}+
2 \alpha_{2}{\frac{\partial \rho_{B}}{\partial t}}+
\beta{\frac{\partial }{\partial t}}(\vec{\nabla}^{2}\rho_{B})
\label{htempcaseIII}
\end{equation}

\section{The KdV equation}

In this section we  repeat the steps developed in \cite{nois,frsw}. We restrict ourselves to 
the one dimensional case ($x$,$t$) and introduce dimensionless variables for the baryon 
density and velocity: 
\begin{equation}
\hat\rho={\frac{\rho_{B}}{\rho_{0}}} \hspace{0.2cm}, \hspace{0.5cm} \hat v={\frac{v}{c_{s}}}
\label{varschapeu}
\end{equation}

We next  define the ``stretched coordinates''  $\xi$ and $\tau$ as in 
\cite{frsw,abu,davidson}:
\begin{equation}
\xi=\sigma^{1/2}{\frac{(x-{c_{s}}t)}{R}} 
\hspace{0.2cm}, \hspace{0.5cm} 
\tau=\sigma^{3/2}{\frac{{c_{s}}t}{R}} 
\label{stret}       
\end{equation}
where $R$ is a size scale and $\sigma$ is a small ($0 < \sigma < 1$) expansion parameter 
chosen to be \cite{davidson}:
\begin{equation}
{\sigma} = {\frac{\mid u-{c_{s}} \mid}{{c_{s}}}}
\label{sigma}       
\end{equation}
where $u$ is the propagation speed of the perturbation in question.  
We then expand (\ref{varschapeu})  around  the equilibrium values:
\begin{equation}
\hat\rho=1+\sigma \rho_{1}+ \sigma^{2} \rho_{2}+ \dots
\label{roexp}
\end{equation}
\begin{equation}
\hat v=\sigma v_{1}+ \sigma^{2} v_{2}+ \dots
\label{vexp}
\end{equation}
After the expansion above  (\ref{conf}) and (\ref{eulerfinalf}) will contain power series in 
$\sigma$   (in practice we go up to $\sigma^2$). Since the coefficients in these series are 
independent of each other we get a set of equations, which, when combined, lead to the KdV 
equation for $\rho_{1}$:
\begin{equation}
{\frac{\partial {\rho}_{1}}{\partial \tau}}+
\bigg({\frac{3}{2}}+{\frac{\Phi{\rho_{0}}^{2}}{2\mu_{B}{c_{s}}^{2}}}-{c_{s}}^{2}\bigg)
{{{\rho}_{1}}{\frac{\partial{\rho}_{1}}{\partial \xi}}}
+\bigg({\frac{\omega\rho_{0}}{2\mu_{B}{c_{s}}^{2}R^{2}}}\bigg)
{\frac{\partial^{3}{\rho}_{1}}{\partial \xi^{3}}}=0  
\label{KdVgeral}
\end{equation}
with the condition
\begin{equation}
{\frac{(\Phi\rho_{0}+\phi)\rho_{0}}{\mu_{B} {c_{s}}^{2}}}=1
\label{condition}
\end{equation}
and where
\begin{equation}
\Phi \equiv \left\{ \begin{array}{ll}
6 \alpha_{3} & \textrm{models  II and III}\\
%6 \alpha_{3} & \textrm{ansatz with saturation}\\
{\frac{3M{c_{s}}^{2}}{{\rho_{0}}^{2}}} & \textrm{model I}
\end{array} \right.
\label{fisão}
\end{equation}
\begin{equation}
\phi \equiv \left\{ \begin{array}{ll}
2 \alpha_{2} & \textrm{model III}\\
-4 \alpha_{3}\rho_{0} & \textrm{model II}\\
{\frac{-2M{c_{s}}^{2}}{{\rho_{0}}}} & \textrm{model I}
\end{array} \right.
\label{fi}
\end{equation}\begin{equation}
\omega \equiv \left\{ \begin{array}{ll}
\beta & \textrm{models II and III}\\
{\frac{{g_{v}}^{2}}{{m_{v}}^{4}}} & \textrm{model I}
\end{array} \right.
\label{omega}
\end{equation}

The equation (\ref{KdVgeral}) has a well known soliton solution. 
We may rewrite the last equation back in  
the $x \, - \, t$ space obtaining a KdV-like equation for ${\hat{\rho}_{1}}$ with the 
following analytical  solitonic solution:
\begin{equation}
{\hat{\rho}_{1}}(x,t)={\frac{3(u-{{c_{s}}})}{{c_{s}}
\bigg({\frac{3}{2}}+{\frac{\Phi{\rho_{0}}^{2}}{2\mu_{B}{c_{S}}^{2}}}-{c_{s}}^{2}\bigg)}}
\,\, sech^{2}\bigg[\sqrt{{\frac{\mu_{B} c_{s}(u-{{c_{s}}})}{2w{\rho_{0}}}}}
(x-ut) \bigg] 
\label{solchapeu}
\end{equation}
where ${\hat{\rho}_{1}} \equiv \sigma {\rho_{1}}$.  This solution is a bump wich propagates 
with speed $u$, without dissipation and  preserving  its  shape.  The expressions given by 
(\ref{KdVgeral}), (\ref{condition}) and (\ref{solchapeu}) depend
on the choices given by (\ref{fisão}), (\ref{fi}) and (\ref{omega}).

In model I (MQHD), the constraint  (\ref{condition}) implies that $\mu_{B}=M$ and
the general equation (\ref{KdVgeral}) becomes:
\begin{equation}
{\frac{\partial {\rho}_{1}}{\partial \tau}}+
(3-{c_{s}}^{2})
{{{\rho}_{1}}{\frac{\partial{\rho}_{1}}{\partial \xi}}}
+\bigg({\frac{{g_{v}}^{2}{\rho_{0}}}{2M{c_{s}}^{2}{m_{v}}^{4}R^{2}}}\bigg)
{\frac{\partial^{3}{\rho}_{1}}{\partial \xi^{3}}}=0  
%+\bigg({\frac{{g_{v}}^{2}\rho_{0}}{2M {c_{s}}^{2} {{m_{v}}^{4}}} R^{2}}}\bigg)
%{\frac{\partial^{3}{\rho}_{1}}{\partial \xi^{3}}}=0  
\label{KdVmqhdrelat}
\end{equation}
with the solution given by:
\begin{equation}
{\hat{\rho}_{1}}(x,t)=\frac{3}{(3-{c_s}^2)} \, {\frac{(u-{c_{s}})}{{c_{s}}}}
\,\, sech^{2}\bigg[
{\frac{{m_{v}}^{2}}{{g_{v}}}}\sqrt{{\frac{(u-{c_{s}}){c_{s}}M}
{2{\rho_{0}}}}}(x-ut) \bigg] 
\label{solmqhdrelat}
\end{equation}
As a consitency check we take the non-relativistic limit, which, in this 
case, means taking a small sound speed $c^2_s \rightarrow 0$. In this limit 
$(3-{c_{s}}^{2})\cong 3$ and (\ref{KdVmqhdrelat}) and (\ref{solmqhdrelat}) 
coincide the results previously obtained  in \cite{nois}:
\begin{equation}
{\frac{\partial {\rho}_{1}}{\partial \tau}}+
3{{\rho}_{1}}{\frac{\partial{\rho}_{1}}{\partial \xi}}
+\bigg({\frac{{g_{v}}^{2}{\rho_{0}}}{2M{c_{s}}^{2}{m_{v}}^{4}R^{2}}}\bigg)
{\frac{\partial^{3}{\rho}_{1}}{\partial \xi^{3}}}=0  
\label{KdVpaper}
\end{equation}
and
\begin{equation}
{\hat{\rho}_{1}}(x,t)={\frac{(u-{c_{s}})}{{c_{s}}}}
\,\, sech^{2}\bigg[
{\frac{{m_{v}}^{2}}{{g_{v}}}}\sqrt{{\frac{(u-{c_{s}}){c_{s}}M}
{2{\rho_{0}}}}}(x-ut) \bigg] 
\label{solpaper}
\end{equation}
In the limit where  $c_s$ is large the factor $3/(3-c^2_s)$ will enhance the soliton 
amplitude with respect to the non-relativistic case. This indicates that in a medium 
with a stiffer EOS the energy propagation through solitary waves is more efficient. 

It is interesting to observe the supersonic nature of the solutions 
(\ref{solmqhdrelat}) and (\ref{solpaper}), which is manifest in the 
arguments of the square roots. As a final remark about (\ref{solchapeu}) 
we notice that, for 
\begin{equation}
{\frac{\Phi{\rho_{0}}^{2}}{2\mu_{B}{c_{S}}^{2}}} < {c_{s}}^{2}-{\frac{3}{2}}
\end{equation}
the solution (\ref{solchapeu}) becomes negative and, in view of (\ref{roexp}),  
can be interpreted as  a rarefaction wave. A solution of this type was found in \cite{abu} 
where nuclear matter was described by an EOS based on the Skyrme force.
\section{Conclusions}

The existence of KdV solitons in nuclear matter has potential applications in 
nuclear physics at intermediate energies \cite{frsw} and also possibly at high 
energies. The experimental measurements of jet quenching and related phenomena 
performed at RHIC \cite{star} offer an unique opportunity of studying supersonic 
motion in hot and dense hadronic matter. With this scenario in mind we took  the 
first steps in the adaptation of the KdV soliton formalism to the new environment. 
We have extended the results of our previous work \cite{nois}, 
showing that it is possible to obtain the KdV solitons in  relativistic  
hydrodynamics.  Moreover we have explored other equations of sate.  
Taking the  non-relativistic limit ($c^2_s \rightarrow 0$)  we were able to  recover the 
previous results.

\begin{acknowledgments}
We wish to express our gratitude to S. Raha for numerous suggestions and useful 
comments and hints. This work was  partially 
financed by the Brazilian funding
agencies CAPES, CNPq and FAPESP. 
\end{acknowledgments}

%%%%%%%%%%%%%%%%%%%%%%%%%%%%%%%%%%%%%%%%%%%%%%%%%%%%%%%%%%%%%%%%%%%%%%%%%%%%%%%

%%%%%%%%%%%%%%%%%%%%%%%%%%%%%% Bibliography %%%%%%%%%%%%%%%%%%%%%%%%%%%%

%\begin{references}

\end{document}